\documentclass[a4paper,fleqn]{cas-sc}

\IfFileExists{charis.sty}{}{\usepackage{lmodern}}

\usepackage[numbers,sort&compress]{natbib}
\usepackage{booktabs}
\usepackage{array}
\usepackage{multirow}
\usepackage{amsmath}
\usepackage{listings}
\usepackage{xcolor}
\usepackage{graphicx}
\usepackage{tikz}
\usetikzlibrary{arrows.meta,positioning,fit,calc,backgrounds}

\definecolor{inkA}{HTML}{0B0B0B}
\definecolor{inkB}{HTML}{52514E}
\definecolor{rule}{HTML}{C3C2B7}
\definecolor{wash}{HTML}{F0EFEC}
\definecolor{cEmb}{HTML}{FCE0E0}   
\definecolor{cAtt}{HTML}{FDE3C3}   
\definecolor{cNorm}{HTML}{F3F4C6}  
\definecolor{cFF}{HTML}{CFE8F6}    
\definecolor{cHead}{HTML}{E0DDF3}  
\definecolor{famTF}{HTML}{0072B2}  
\definecolor{famEnc}{HTML}{009E73} 
\definecolor{famZS}{HTML}{E69F00}  
\definecolor{famFT}{HTML}{D55E00}  

\ExplSyntaxOn
\RenewDocumentCommand \printorcid { }
  {
    \seq_if_empty:NF \g_stm_orcid_seq
      {
        \group_begin:
          \tex_let:D \thefootnote \relax
          \footnotetext
            { \raggedright \textsc{orcid}(s):\c_space_token \seq_use:Nn \g_stm_orcid_seq { ;~ } }
        \group_end:
      }
  }
\ExplSyntaxOff

\ExplSyntaxOn
\cs_set:Npn \__first_footerline:
  { \group_begin: \small \sffamily \__short_authors: :~ { \rmfamily \itshape Preprint } \group_end: }
\ExplSyntaxOff

\newcommand{\coderepo}{\url{https://github.com/tbayan/bug_finetuning_llm_artifacts}}

\newif\ifjournalversion
\journalversionfalse

\AtBeginDocument{%
  \setcounter{topnumber}{3}\setcounter{bottomnumber}{2}\setcounter{totalnumber}{5}%
  }

\begin{document}
\let\WriteBookmarks\relax

\shorttitle{Fine-tuned LLMs for bug severity classification}
\shortauthors{A. Kanatov et~al.}

\title[mode=title]{Bug Severity Classification with Fine-Tuned Open LLMs: A Paired
Comparison with TF-IDF and CodeBERT Baselines on Eclipse Reports}

\author[1]{Asset Kanatov}
\ead{asset.kanatov@nu.edu.kz}
\credit{Conceptualization, Methodology, Software, Investigation, Data curation,
Writing -- original draft}

\affiliation[1]{organization={School of Computing and Artificial Intelligence (SCAI), Nazarbayev University},
                city={Astana},
                country={Kazakhstan}}

\author[1,2]{Talgar Bayan}
\ead{talgar.bayan@nu.edu.kz}
\credit{Methodology, Supervision, Validation, Formal analysis, Writing -- review \& editing}

\affiliation[2]{organization={University of Manchester},
                city={Manchester},
                country={United Kingdom}}

\author[1]{Askar Boranbayev}
\cormark[1]
\ead{askar.boranbayev@nu.edu.kz}
\credit{Supervision, Writing -- review \& editing}

\cortext[1]{Corresponding author}

\begin{abstract}
Open large language models (LLMs) small enough to run on a laptop are an obvious
candidate for automatic bug severity prediction. We measure what they add over cheaper
classifiers on one task: assigning one of four Bugzilla severity levels (Critical,
Major, Minor, Trivial) to Eclipse bug reports. From a public dump of 361,006 unique
Eclipse reports we draw a class-balanced sample of 20,000 rows and compare eleven
configurations of nine model types: six TF-IDF classifiers, a fine-tuned CodeBERT
encoder, and Llama~3~8B and Qwen~2.5~7B, each run zero-shot and after QLoRA fine-tuning.
All eleven are scored on the same 2,000 held-out reports, with bootstrap intervals and
McNemar tests. The zero-shot runs reach macro-F1 0.305 and 0.321, below every TF-IDF
model (0.430--0.489). The fine-tuned runs reach 0.526 and 0.485, but the zero-shot and
fine-tuned setups also differ in checkpoint, serving runtime and decoding, so this gap
cannot be credited to fine-tuning alone. Fine-tuned Llama~3 shows no detectable
difference from CodeBERT (0.518; difference $+0.007$, 95\% CI $-0.017$ to $+0.032$),
and fine-tuned Qwen~2.5 is below it on macro-F1. The models differ more by class than in aggregate:
fine-tuned Llama~3 finds 36\% of Critical reports, Qwen~2.5 finds 69\% but only 21\% of
Minor ones. The dump repeats 60,000 rows, and 8\% of the test reports duplicate a
training report; removing them lowers TF-IDF scores by up to 0.031 and leaves the LLMs
almost unchanged. We release split identifiers, per-report predictions and the analysis
script at \coderepo, together with CyberTriage, a tool that serves the fine-tuned classifier locally.
\end{abstract}


\maketitle

\section{Introduction}
\label{sec:intro}

When a reporter files a bug in Bugzilla, they pick a severity from a fixed list that
runs from \textit{blocker} to \textit{trivial}, and triagers use that field to decide
what gets looked at first. A wrong value costs most when the defect is a security or
privacy weakness. Such weaknesses keep reaching deployed software: in permissionless
blockchains, privacy flaws~\citep{bayan2023exploring} and smart-contract and signature
vulnerabilities have led to documented incidents, and a secure development lifecycle
with measurable security controls has been proposed to catch them during
development~\citep{bayan2025permissionless}. The field is often
unreliable. In the Eclipse collection we
study, two thirds of all reports are labelled \textit{normal}, the default value, and
\citet{tian2016unreliability} found that about half of the duplicate reports in
Eclipse, Mozilla and OpenOffice carry different severities even though they describe
the same problem. Predicting severity from the report text has therefore been studied
for well over a decade, starting with rule learners and Naive Bayes over word
features~\citep{menzies2008automated,lamkanfi2010predicting} and moving through deep
networks~\citep{ramay2019deep,dao2021severity} to pre-trained
encoders~\citep{ali2024bert,agarwal2026finetuning} and, recently, generative
LLMs~\citep{oliveira2024predictive,arshad2024sevpredict,ali2025retrieval}.

Open models with 7--8 billion parameters are a natural next step. QLoRA lets them be
adapted on a single GPU~\citep{dettmers2023qlora}, and once quantised they run on a
laptop, so a team could keep its reports in-house. What is less clear is whether they
beat a fine-tuned encoder that is sixty times smaller. Published comparisons rarely
settle this, because they differ in data, label scheme, split and metric, and few of
them score every model on the same reports with a measure of uncertainty.

This paper compares eleven configurations of nine model types on one task and one
split: TF-IDF classifiers, a fine-tuned CodeBERT encoder~\citep{feng2020codebert}, and
two open LLMs, Llama~3~8B~\citep{grattafiori2024llama} and
Qwen~2.5~7B~\citep{qwen2024qwen25}, used both zero-shot and after QLoRA fine-tuning.
All are scored on the same 2,000 test reports. The work grew out of a master's thesis.
The LLM scores reported in that thesis, and those shown by the evaluation screen of the
original CyberTriage code, were computed by a scoring routine that added synthetic
correct predictions, 25\% as many as the real ones, before computing the metrics. For
this paper we recomputed every number with one script from the raw data dump, the real
per-report predictions and the training logs; these numbers supersede the thesis
values, and we report only what the records support. The code released with this paper
no longer uses that routine and keeps it only as a record. As in our earlier data
releases, which publish the raw outputs and aggregated metrics of every
run~\citep{bayan2026new,bayan2026dataset}, we release the split identifiers, the
per-report predictions and the analysis script. Three questions guide the analysis.

\noindent\textbf{RQ1.} On identical test reports, how do QLoRA-fine-tuned 7--8B LLMs
compare with TF-IDF classifiers and a fine-tuned CodeBERT encoder?

\noindent\textbf{RQ2.} What changes when the same two model families are fine-tuned
instead of prompted zero-shot, overall and for each severity level?

\noindent\textbf{RQ3.} How sensitive are these comparisons to two properties of the
data: reports duplicated across the split, and the balanced class distribution?

\noindent The paper makes four contributions:
\begin{enumerate}
\item A paired comparison of eleven configurations on the same 2,000 reports, with
bootstrap intervals, McNemar tests and Holm correction, released with split
identifiers and per-report predictions.
\item An audit of the public Hugging Face dump used here, showing that it holds Eclipse
reports only, repeats 60,000 rows, has fewer unique Trivial reports than the sampling
asked for, and lets 8\% of the test reports duplicate a training report, together with
the measured effect of each issue.
\item Measurements of the two fine-tuned LLMs in this setting: far above the zero-shot
runs, though in a setup that differs in more than training; no detectable difference
from CodeBERT at about 27 times its per-report latency; and different per-class
trade-offs between the two.
\item CyberTriage, an open web tool that serves the fine-tuned classifier on a laptop
and adds a rationale written by a second LLM, which cannot change the stored label.
\end{enumerate}

Section~\ref{sec:related} reviews severity prediction. Section~\ref{sec:data}
describes the data and what the audit found, and Section~\ref{sec:method} describes the models
and evaluation protocol. Section~\ref{sec:results} answers the three questions,
Section~\ref{sec:system} describes CyberTriage, Section~\ref{sec:discussion} discusses
the findings and threats to validity, and Section~\ref{sec:conclusion} concludes.

\section{Related work}
\label{sec:related}

\subsection{Severity prediction from report text}

\citet{menzies2008automated} were among the first to predict severity automatically,
applying the Ripper rule learner to terms selected with TF-IDF and information gain from
NASA defect reports with five severity levels. \citet{lamkanfi2010predicting} framed the
Bugzilla version of the task as a binary one, grouping \textit{blocker},
\textit{critical} and \textit{major} as severe and \textit{minor} and \textit{trivial}
as non-severe, and trained Naive Bayes on three open-source projects; a follow-up
comparison on Eclipse and GNOME found multinomial Naive Bayes ahead of plain Naive
Bayes, $k$-nearest neighbours and support vector machines~\citep{lamkanfi2011comparing}.
\citet{tian2012information} kept the five Bugzilla levels and predicted them with a
BM25-based nearest-neighbour method whose similarity weights were learned from
duplicate reports, and \citet{yang2014towards} combined topic models with several report
features to find similar historical reports for severity prediction. Later work
replaced hand-built features with neural networks: \citet{ramay2019deep} fed text
vectors and an emotion score to a deep network, and \citet{dao2021severity} combined
content, sentiment, quality and reporter features in a convolutional network evaluated
on Eclipse and Mozilla.

Two points from this work matter here. Many studies drop the
\textit{normal} label, because it is the Bugzilla default and many reporters never
change it~\citep{oliveira2024predictive}, and the label itself is noisy:
\citet{tian2016unreliability} found inconsistent severities for about 51\% of duplicate
reports and argued that agreement with the recorded label is an imperfect yardstick.

\subsection{Pre-trained encoders and generative models}

Pre-trained transformers entered the task as encoders. \citet{ali2024bert} fine-tuned
BERT together with a sentiment signal on mobile-application reports, and
\citet{agarwal2026finetuning} compared fine-tuned code models for severity prediction
in a study built on GraphCodeBERT and LoRA. Generative models followed. SevPredict fine-tunes
GPT-2 and reports gains over a BERT-based predictor~\citep{arshad2024sevpredict};
RAG-GPT-SBR adds retrieved historical reports and metadata to a fine-tuned GPT-2 with a
classification head~\citep{ali2025retrieval}; and \citet{oliveira2024predictive}
prompted and fine-tuned Llama~2 for the severe versus non-severe split on Eclipse and
Mozilla. Table~\ref{tab:related} summarises these studies. Their scores are not
comparable with each other or with ours, since data, label schemes and splits differ.

Beyond severity, \citet{hou2024large} survey LLM use across software engineering, and
\citet{kiashemshaki2025automated} apply instruction-tuned LLMs with constrained decoding
to developer assignment, a different triage task. LoRA~\citep{hu2022lora} trains
low-rank updates next to frozen weights, and QLoRA~\citep{dettmers2023qlora} keeps the
frozen weights in 4-bit NormalFloat (NF4) so that a 7--8B model can be tuned on one GPU.
We use QLoRA as an off-the-shelf method; it is not a contribution of this paper.

\begin{table*}[width=\textwidth,cols=4,pos=t]
\caption{Selected severity-prediction studies from Section~\ref{sec:related}. Label
schemes and data differ, so reported scores cannot be compared across rows. A dash
marks information we could not confirm from the published abstract or record.}
\label{tab:related}
\small
\begin{tabular*}{\tblwidth}{@{\extracolsep{\fill}} p{3.6cm} p{3.6cm} p{3.4cm} p{4.6cm} @{}}
\toprule
Study & Data & Labels & Models \\
\midrule
\citet{menzies2008automated} & NASA PITS & five NASA levels & Ripper rule learner \\
\citet{lamkanfi2010predicting} & three open-source projects & severe vs.\ non-severe & Naive Bayes \\
\citet{lamkanfi2011comparing} & Eclipse, GNOME & severe vs.\ non-severe & NB, multinomial NB, $k$-NN, SVM \\
\citet{tian2012information} & three open-source projects & blocker to trivial & BM25 nearest neighbour \\
\citet{dao2021severity} & Eclipse, Mozilla & --- & CNN with multi-aspect features \\
\citet{oliveira2024predictive} & Eclipse, Mozilla & severe vs.\ non-severe & Llama~2 (prompted, fine-tuned) \\
\citet{agarwal2026finetuning} & --- & --- & GraphCodeBERT and other code models, LoRA \\
\citet{ali2025retrieval} & mobile-app reports & --- & fine-tuned GPT-2 with retrieval \\
This study & Eclipse & four levels & TF-IDF, CodeBERT, Llama~3, Qwen~2.5 \\
\bottomrule
\end{tabular*}
\end{table*}

\section{Data}
\label{sec:data}

\subsection{Source and provenance}

We use the Hugging Face dataset \texttt{AndressaStefany/bug-reports}\footnote{\url{https://huggingface.co/datasets/AndressaStefany/bug-reports}}, a CSV export with six columns: an internal record identifier, the Bugzilla
\texttt{bug\_id}, the free-text \texttt{description}, \texttt{bug\_severity},
\texttt{bug\_status} and \texttt{resolution}. The dataset has no card, so we identified
it from its contents. The file holds 421,006 rows, but the last 60,000 rows are exact copies of
earlier ones (bug IDs 1 to 62,767), leaving 361,006 unique reports with bug IDs up to
424,785. The repeats come from the dataset repository itself: it stores two CSV files,
\texttt{eclipse\_clear.csv} with 361,006 rows and \texttt{my\_dataset.csv} with 60,000
rows that copy the first 60,000 of the former, and the standard loader joins both into
one split. That count and the per-label counts in Table~\ref{tab:dataset} match exactly
the Eclipse collection described by \citet{oliveira2024predictive}, who also used a
separate Mozilla collection of 768,335 reports that is not part of this dump. The
resolution values include \texttt{NOT\_ECLIPSE}, and 38\% of the unique descriptions
mention Eclipse by name. We therefore treat the data as Eclipse-only.

Each report is represented by its description alone; the dump has no summary line.
Every report in it is resolved, closed or verified, but status and resolution were not
given to any model. Some descriptions, mostly from the earliest years, carry text added
after filing, such as a \texttt{NOTES:} section or dated entries signed with initials;
201 of the 20,000 sampled descriptions (1.0\%) contain such markers. The severity is the value of the
field when the data were collected; the dump keeps no history, so we cannot tell
initial labels from revised ones.

\subsection{Labels and exclusions}

Table~\ref{tab:dataset} lists every label in the dump. Two of them were excluded:
\textit{normal}, which covers 67.3\% of unique reports and is the Bugzilla default, and
\textit{enhancement}, which marks feature requests (12.6\%). The remaining five were
mapped to four classes by merging \textit{blocker} and \textit{critical} into Critical.
After the mapping, 432 rows with empty descriptions were dropped, leaving a pool of
83,488 rows (72,337 unique reports). The pool is far from balanced and is dominated by
the more severe classes: among unique reports, Major makes up 48.6\%, Critical 26.3\%,
Minor 18.5\% and Trivial 6.6\%.

\begin{table}[width=\linewidth,cols=6,pos=t]
\caption{Labels in the Hugging Face dump and how the experimental sample was drawn.
``Rows'' counts every row in the file; ``Unique reports'' drops the 60,000 repeated
rows. The pool is what the sampler drew from; it still contains repeated rows.}
\label{tab:dataset}
\small
\begin{tabular*}{\tblwidth}{@{\extracolsep{\fill}} lrrlrr @{}}
\toprule
Label in the dump & Rows & Unique reports & Mapped class & Pool rows & Sampled (train / test) \\
\midrule
\texttt{blocker} & 6,972 & 6,223 & Critical & 22,127 & 4,000 / 1,000 \\
\texttt{critical} & 15,255 & 12,845 &  &  &  \\
\texttt{major} & 40,743 & 35,367 & Major & 40,530 & 4,000 / 1,000 \\
\texttt{minor} & 15,539 & 13,438 & Minor & 15,461 & 4,000 / 1,000 \\
\texttt{trivial} & 5,411 & 4,815 & Trivial & 5,370 & 4,000 / 1,000 \\
\texttt{normal} & 282,829 & 242,819 & excluded & --- & --- \\
\texttt{enhancement} & 54,257 & 45,499 & excluded & --- & --- \\
\midrule
Total & 421,006 & 361,006 & & 83,488 & 16,000 / 4,000 \\
\bottomrule
\end{tabular*}

\end{table}

\subsection{Sampling, split and what the split contains}

The preparation script (\texttt{dataset/1\_prep\_balance\_split\_csv.py}) drew 5,000
rows per class without replacement from the pool with \texttt{random\_state=42} and
split the 20,000 rows 80/20, stratified by class, into 16,000 training and 4,000 test
reports. Rerunning the script on the dump reproduces both files exactly, which lets us
attach a bug ID to every row; the identifiers are released with this paper.

Sampling from rows rather than unique reports has two consequences. First, the pool
holds only 4,778 unique Trivial reports, fewer than the 5,000 requested, so repeated
rows were bound to enter the sample; 962 bug IDs appear twice in the 20,000 rows, 519
of them Trivial, so the sample holds 19,038 distinct reports. Second, repeats can land on both sides of the split: 319
test reports (8.0\%) have a description identical to a training report, 175 of them
Trivial. Balancing before splitting does not by itself prevent this kind of leakage.
Section~\ref{sec:res_sensitivity} measures its effect. The text itself also shows some
label noise: among the unique pool reports, 41 descriptions are filed under more than
one bug ID with conflicting severities (243 reports in all). We did not inspect a
sample of reports manually, so we make no claim about the overall noise rate.

The sample mixes outcomes. Only 64.4\% of the 20,000 reports were resolved as
\texttt{FIXED}; the rest were closed as duplicates (2,673, counting the
\texttt{NDUPLICATE} value), works-for-me (1,563), invalid (1,394) or won't-fix (1,267),
among others. Descriptions have a median length of 413 characters, and 6.4\% are longer
than the 4,000-character limit applied to LLM inputs.

Both training and test sets are balanced. The study has no test set at natural
prevalence; Section~\ref{sec:res_sensitivity} estimates how the scores would shift
under the pool's class mix, and Section~\ref{sec:threats} discusses what that estimate
cannot cover.

\section{Models and evaluation protocol}
\label{sec:method}

Figure~\ref{fig:design} shows the study design and Figure~\ref{fig:arch} the two neural
architectures that were trained. Each model family is described below with the
settings recovered from the code and logs; Table~\ref{tab:training} collects the
training configurations of the learned models.

\begin{figure}[pos=t]
\centering
\begin{tikzpicture}[font=\sffamily\footnotesize, >={Stealth[length=1.8mm]},
  data/.style={draw=black, fill=gray!12, line width=0.5pt, align=center, inner sep=4pt, text=inkA, minimum height=11mm},
  model/.style={draw=black, fill=white, line width=0.5pt, align=center, inner sep=3pt, text=inkA, minimum width=5.1cm},
  arr/.style={->, draw=black, line width=0.5pt}]
  \node[data] (dump) {Hugging Face dump\\421,006 rows, of which 60,000\\are exact repeats};
  \node[data, right=5mm of dump] (pool) {drop \textit{normal}, \textit{enhancement};\\map 5 labels to 4\\pool: 83,488 rows};
  \node[data, right=5mm of pool] (samp) {5,000 rows per class\\20,000 rows\\(19,038 distinct reports)};
  \node[data, below=8mm of samp.south east, anchor=north east] (test) {test: 4,000\\319 texts also\\in training};
  \node[data, left=4mm of test] (train) {train:\\16,000};
  \node[data, below=6mm of test] (shared) {shared subset\\2,000 reports};
  \draw[arr] (dump) -- (pool);
  \draw[arr] (pool) -- (samp);
  \draw[arr] (samp.south) -- ++(0,-3mm) -| (train.north);
  \draw[arr] (samp.south) -- ++(0,-3mm) -| (test.north);
  \draw[arr] (test) -- (shared);
  \node[model, fill=famTF!15, below=8mm of dump.south west, anchor=north west] (tf) {6 TF-IDF classifiers};
  \node[model, fill=famEnc!15, below=1.5mm of tf] (enc) {CodeBERT, full fine-tuning};
  \node[model, fill=famFT!15, below=1.5mm of enc] (ft) {Llama 3 8B, Qwen 2.5 7B, QLoRA};
  \node[model, fill=famZS!18, below=1.5mm of ft] (zs) {Llama 3 8B, Qwen 2.5 7B, zero-shot};
  \draw[arr] (train.west) -- ++(-3mm,0) |- (tf.east);
  \draw[arr] (train.west) -- ++(-3mm,0) |- (enc.east);
  \draw[arr] (train.west) -- ++(-3mm,0) |- (ft.east);
  \path let \p1 = (zs.south west), \p2 = (shared.east) in
    node[draw=black, line width=0.5pt, align=center, inner sep=4pt, text=inkA, anchor=north west,
         minimum width=\x2-\x1] (eval) at ([yshift=-6mm]zs.south west |- shared.south)
    {all 11 configurations on the same 2,000 reports: macro-F1 with bootstrap intervals,\\
     McNemar tests with Holm correction; sensitivity to repeated reports and class mix};
  \coordinate (bus) at ([xshift=-3mm]tf.west);
  \foreach \n in {tf, enc, ft, zs} \draw[line width=0.5pt] (\n.west) -- (\n.west -| bus);
  \draw[arr] (tf.west -| bus) |- (eval.west);
  \draw[arr] (shared.south) -- (shared.south |- eval.north);
\end{tikzpicture}
\caption{Study design. Grey boxes are data, with the counts found in the audit; coloured
boxes are the eleven model configurations, in the family colours of Figure~\ref{fig:f1}. The TF-IDF classifiers were fit on 12,800 of
the 16,000 training reports. The shared subset is the 2,000-report sample on which the
LLMs were run; every other model was re-scored on it from its stored files.}
\label{fig:design}
\end{figure}
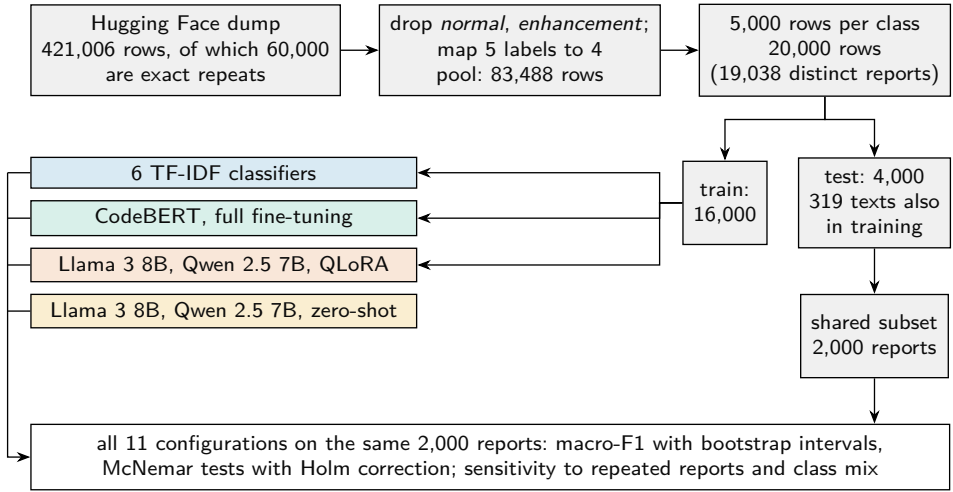

\begin{figure*}[pos=t]
\centering
\begin{tikzpicture}[font=\sffamily\scriptsize, >={Stealth[length=1.6mm]},
  node distance=2.6mm,
  box/.style={draw=black, fill=white, rounded corners=1.5pt, line width=0.5pt,
              minimum height=5.4mm, align=center, text=inkA},
  enc/.style={box, minimum width=3.2cm},
  frz/.style={box, minimum width=3.0cm},
  emb/.style={fill=cEmb}, att/.style={fill=cAtt}, nrm/.style={fill=cNorm},
  ffn/.style={fill=cFF}, hd/.style={fill=cHead}, lora/.style={fill=famFT!25},
  io/.style={align=center, text=inkA},
  add/.style={circle, draw=black, line width=0.5pt, inner sep=0pt, minimum size=3.4mm, font=\sffamily\footnotesize},
  tag/.style={draw=black, lora, rounded corners=1pt, inner sep=1.4pt, font=\sffamily\tiny, text=inkA},
  arr/.style={->, draw=black, line width=0.5pt},
  blk/.style={draw=black, line width=0.5pt, fill=gray!7, rounded corners=4pt, inner sep=2.4mm},
  ttl/.style={font=\sffamily\footnotesize\bfseries, text=inkA}]

\node[io] (ein) {report text\\(first 512 tokens)};
\node[enc, emb, above=3.5mm of ein] (eemb) {token + position embedding,\\LayerNorm};
\node[enc, att, above=6mm of eemb] (eatt) {multi-head self-attention\\(12 heads)};
\node[enc, nrm, above=of eatt] (ean1) {add \& LayerNorm};
\node[enc, ffn, above=of ean1] (eff) {feed-forward\\768 $\to$ 3,072 $\to$ 768, GELU};
\node[enc, nrm, above=of eff] (ean2) {add \& LayerNorm};
\begin{scope}[on background layer]
  \node[blk, fit=(eatt)(ean2)(eff), inner xsep=6mm] (eblk) {};
\end{scope}
\node[left=1mm of eblk, font=\sffamily\small, text=inkA] {12$\times$};
\node[enc, above=6mm of ean2] (ecls) {first-token vector (768)};
\node[enc, hd, above=of ecls] (ehead) {classification head:\\dense, tanh, linear 768 $\to$ 4};
\node[io, above=3.5mm of ehead] (eout) {softmax over the\\four severity labels};
\foreach \a/\b in {ein/eemb, eemb/eatt, eatt/ean1, ean1/eff, eff/ean2, ean2/ecls, ecls/ehead, ehead/eout}
  \draw[arr] (\a) -- (\b);
\coordinate (er1) at ($(eemb.north)!0.5!(eatt.south)$);
\draw[arr, rounded corners=2pt] (er1) -- ([xshift=1.8mm]eatt.east |- er1) |- (ean1.east);
\coordinate (er2) at ($(ean1.north)!0.5!(eff.south)$);
\draw[arr, rounded corners=2pt] (er2) -- ([xshift=3.6mm]eff.east |- er2) |- (ean2.east);

\node[io, right=3.1cm of ein] (din) {prompt of Listing~1:\\instruction + report\\(first 4,000 characters)};
\node[frz, emb, above=3.5mm of din] (demb) {token embedding};
\node[frz, nrm, above=6mm of demb] (dn1) {RMSNorm};
\node[frz, att, above=of dn1] (datt) {grouped-query\\self-attention, RoPE\\projections q, k, v, o};
\node[add, above=2mm of datt] (da1) {$+$};
\node[frz, nrm, above=2mm of da1] (dn2) {RMSNorm};
\node[frz, ffn, above=of dn2] (dff) {SwiGLU feed-forward\\projections gate, up, down};
\node[add, above=2mm of dff] (da2) {$+$};
\begin{scope}[on background layer]
  \node[blk, fit=(dn1)(da2)(datt)(dff), inner xsep=7mm] (dblk) {};
\end{scope}
\node[left=1mm of dblk, font=\sffamily\small, text=inkA] {$N\times$};
\node[tag, anchor=west] at ([xshift=0.8mm]datt.east) {+LoRA};
\node[tag, anchor=west] at ([xshift=0.8mm]dff.east) {+LoRA};
\node[frz, nrm, above=6mm of da2] (dnf) {RMSNorm};
\node[frz, hd, above=of dnf] (dlm) {LM head $\to$ next-token\\probabilities};
\node[io, above=3.5mm of dlm] (dout) {generated answer, e.g.\ \texttt{\{"severity": "Major"\}},\\parsed into the label (Section~\ref{sec:protocol})};
\foreach \a/\b in {din/demb, demb/dn1, dn1/datt, datt/da1, da1/dn2, dn2/dff, dff/da2, da2/dnf, dnf/dlm, dlm/dout}
  \draw[arr] (\a) -- (\b);
\coordinate (dr1) at ($(demb.north)!0.5!(dn1.south)$);
\draw[arr, rounded corners=2pt] (dr1) -- ([xshift=-2mm]datt.west |- dr1) |- (da1.west);
\coordinate (dr2) at ($(da1.north)!0.5!(dn2.south)$);
\draw[arr, rounded corners=2pt] (dr2) -- ([xshift=-4mm]datt.west |- dr2) |- (da2.west);

\node[io, below=5mm of $(ein)!0.5!(din)$, draw=rule, fill=wash, rounded corners=1.5pt, inner sep=3pt] (rep) {bug report description};
\draw[arr] (rep.west) -| (ein.south);
\draw[arr] (rep.east) -| (din.south);
\coordinate (top) at ([yshift=3.5mm]dout.north);
\node[ttl, anchor=south] at (eout |- top) {(a) CodeBERT, full fine-tuning};
\node[ttl, anchor=south] at (dout |- top) {(b) Llama 3 8B, Qwen 2.5 7B with QLoRA};

\begin{scope}[shift={($(dblk.east |- dn2)+(2.15cm,-0.2cm)$)}]
  \node[io] (lx) at (0,0) {$x \in \mathbb{R}^{k}$};
  \node[frz, minimum width=1.55cm, minimum height=1.2cm] (lw) at (-0.95,1.45) {$W_0$\\$d \times k$\\frozen, NF4};
  \node[box, lora, minimum width=1.25cm, minimum height=4.6mm] (la) at (0.95,0.95) {$A$: $r \times k$};
  \node[box, lora, minimum width=1.25cm, minimum height=4.6mm] (lb) at (0.95,1.95) {$B$: $d \times r$};
  \node[add] (lsum) at (0,2.95) {$+$};
  \node[io] (lh) at (0,3.55) {$h = W_0x + \tfrac{\alpha}{r}BAx$};
  \draw[arr] (lx.north) -- ++(0,0.22) -| (lw.south);
  \draw[arr] (lx.north) -- ++(0,0.22) -| (la.south);
  \draw[arr] (la) -- (lb);
  \draw[arr] (lw.north) |- (lsum.west);
  \draw[arr] (lb.north) |- (lsum.east);
  \node[font=\sffamily\tiny, text=inkB, anchor=west] at ([xshift=0.5mm, yshift=1.6mm]lb.north) {scale $\alpha/r$};
  \node[ttl, anchor=south] at ([yshift=1mm]lh.north) {(c) one adapted projection};
\end{scope}
\node[io, anchor=north, font=\sffamily\scriptsize] at ([yshift=-3mm]lx.south) {%
  \setlength{\tabcolsep}{2.5pt}%
  \begin{tabular}{@{}lrr@{}}
    & Llama & Qwen \\[-1pt]
    & 3 8B & 2.5 7B \\
    \hline
    blocks $N$ & 32 & 28 \\
    width $d_{\text{model}}$ & 4,096 & 3,584 \\
    feed-forward & 14,336 & 18,944 \\
    heads, q / kv & 32 / 8 & 28 / 4 \\
    LoRA $r$, $\alpha$ & 32, 32 & 32, 32 \\
    trainable & 83.9M & 80.7M
  \end{tabular}};

\node[anchor=north, font=\sffamily\scriptsize, text=inkA] at ([yshift=-2.5mm]rep.south) {%
  \tikz[baseline=0.1ex]{\draw[draw=black, fill=famFT!25, line width=0.5pt, rounded corners=1pt] (0,0) rectangle (4.5mm,2.4mm);}~trained LoRA adapter;
  (a) trains all weights, (b) only the adapters};
\end{tikzpicture}
\caption{Architecture of the two trained neural classifiers. (a) CodeBERT, a 12-layer
RoBERTa encoder fine-tuned end-to-end with a classification head. (b) The decoder-only
LLMs, whose pre-trained weights stay frozen while LoRA adapters are trained on all seven
linear projections of every block; the label is read from the generated text. During
training the block projections are stored in NF4 with double quantisation, while the
embeddings, normalisation layers and LM head stay in 16-bit. (c) One adapted projection
(Eq.~\ref{eq:lora}). Dimensions are taken from the CodeBERT configuration file and the
headers of the deployed GGUF files, and trainable-parameter counts from the training
logs.}
\label{fig:arch}
\end{figure*}

\subsection{TF-IDF classifiers}

Text was vectorised with TF-IDF over word 1--3-grams (20,000 features, sublinear term
frequency, English stop words removed, minimum document frequency 5, maximum 0.75).
Six scikit-learn classifiers~\citep{pedregosa2011scikit} were trained: multinomial
Naive Bayes with uniform class priors, logistic regression, linear SVM, random forest,
XGBoost, and a soft-voting ensemble of logistic regression, a calibrated linear SVM,
random forest and Naive Bayes. Hyperparameters were chosen by 10-fold grid search on
macro-F1 (selected values: $C{=}1$ for logistic regression, $C{=}0.1$ for the SVM, 500
trees and \texttt{min\_samples\_split}${=}5$ for the forest, $\alpha{=}10$ for Naive
Bayes, 200 trees of depth 10 for XGBoost). Inside the ensemble, the members keep fixed
settings ($C{=}1$ for both linear models, 300 trees, $\alpha{=}0.1$) and only the voting
weights were searched, giving $(2,2,1,1)$. Two details of the training
script matter. The classifiers were fit on 12,800 of the 16,000 training reports; the
other 3,200 served only as a printed check. And the vectoriser was fit once on those
12,800 reports before the grid search, so the cross-validation folds share its
vocabulary and inverse document frequencies. This touches hyperparameter choice only;
the test split was not involved.

\subsection{CodeBERT}

CodeBERT~\citep{feng2020codebert} (RoBERTa-base architecture, 124.6M parameters) was
fine-tuned end-to-end with a four-way classification head using Hugging Face
Transformers~\citep{wolf2020transformers}. Inputs were truncated at 512 tokens, which
cuts 13.2\% of the test reports. Training ran for three epochs (Table~\ref{tab:training}).
The training script evaluated on the test split after every epoch and kept the
checkpoint with the best test macro-F1. The three per-epoch values were 0.478, 0.526
and 0.532, so the kept checkpoint is the final one and equals what a fixed three-epoch
schedule would have produced; the test split was nonetheless consulted during
training, and we flag it here rather than retrain.

\subsection{Zero-shot LLMs}

Llama~3~8B and Qwen~2.5~7B were queried through Ollama\footnote{\url{https://github.com/ollama/ollama}}
with the default tags \texttt{llama3:8b} and \texttt{qwen2.5:7b}, which at the time of
writing resolve to the instruction-tuned variants (4-bit Q4\_0 and Q4\_K\_M builds
respectively). The
instruction in Listing~\ref{lst:prompt} was wrapped in the Alpaca layout shown and sent
to Ollama's \texttt{generate} endpoint, which places it in each model's chat template.
Decoding used temperature 0.1 and at most 150 new tokens. No examples were included in
the prompt and no output constraint was imposed; constrained and few-shot variants were
not tested. The Ollama version and the model digests used at run time were not recorded,
and the scripts did not set Ollama's context length, so the installed default applied.

\begin{lstlisting}[float=tb, caption={Instruction and layout used for zero-shot prompting and,
with the gold answer appended, for every fine-tuning record. The fine-tuned models
receive the same text at inference and are expected to complete the response.},
label={lst:prompt}]
Below is an instruction that describes a task, paired with an input that provides further context. Write a response that appropriately completes the request.

### Instruction:
You are a Senior QA Engineer. Classify the severity of the following bug report.

Severity definitions:
- Critical: System crash, data loss, security vulnerability.
- Major: Significant functionality is broken.
- Minor: Small functional issue or inconvenience.
- Trivial: Cosmetic issue or typo.

Output ONLY a valid JSON object with a single key:
{"severity": "<label>"}

### Input:
<bug report description, first 4,000 characters>

### Response:
{"severity": "Major"}<end-of-sequence>        (training records only)
\end{lstlisting}

\subsection{QLoRA fine-tuning}
\label{sec:qlora}

Both LLMs were fine-tuned with Unsloth\footnote{\url{https://github.com/unslothai/unsloth}}
on one NVIDIA A100 (40\,GB) in Google Colab. Training started from the pre-trained base
checkpoints, not the instruction-tuned ones, loaded in NF4. Each of the 16,000
training reports became one record in the format of Listing~\ref{lst:prompt}, so the
fine-tuning prompt carries the same severity definitions as the zero-shot prompt. LoRA
adapters~\citep{hu2022lora} were attached to all seven linear projections of every
transformer block (query, key, value and output in attention; gate, up and down in the
feed-forward network). For a frozen matrix $W_0\in\mathbb{R}^{d\times k}$ the adapted
layer computes
\begin{equation}
  h = W_0 x + \frac{\alpha}{r}\, B A x,\qquad B\in\mathbb{R}^{d\times r},\;
  A\in\mathbb{R}^{r\times k},
  \label{eq:lora}
\end{equation}
with $r=\alpha=32$, so the scaling factor is one; during training, dropout with
probability 0.05 is applied to the adapter input. Each adapted matrix adds $r(d+k)$
trainable parameters. For Llama~3~8B the seven projections in one block contribute
$32\times81{,}920=2{,}621{,}440$ parameters, and the 32 blocks give the 83,886,080
reported in the training log, 1.03\% of the adapted model. For a single
$4096\times4096$ projection the adapter is $dk/\big(r(d+k)\big)=64$ times smaller than
the matrix. That ratio describes what is stored and updated, not compute or memory: the
forward and backward passes still run through the whole frozen network.

Training minimised the next-token cross-entropy computed from the output logits,
averaged over all predicted tokens of an effective batch $\mathcal{B}$ of eight records,
\begin{equation}
  \mathcal{L}(\theta) = -\frac{\sum_{i\in\mathcal{B}}\sum_{t=2}^{T_i}
  \log p_\theta\!\left(x_{i,t} \mid x_{i,<t}\right)}{\sum_{i\in\mathcal{B}}(T_i-1)},
  \label{eq:loss}
\end{equation}
where $x_{i,1},\dots,x_{i,T_i}$ are the tokens of record $i$: instruction, report, answer
and the end-of-sequence token. The first token is never predicted; for Llama~3 it is the
beginning-of-sequence token, which the Qwen~2.5 base tokenizer does not add. The script
did not mask the prompt tokens, so the loss covers the report text as well as the
answer, which is only 6 or 7 tokens long under either tokenizer. All settings are listed
in Table~\ref{tab:training}. No validation split was carved out and no hyperparameter
search was logged. Besides the two runs in Table~\ref{tab:training}, the repository holds
the log of one more Llama~3 run, trained with the same LoRA settings on the unquantised
16-bit base model (paged 32-bit AdamW, 2 h 22 min) and exported as a 16-bit GGUF file; it
was never evaluated and no predictions exist for it. We cannot rule out that earlier,
unrecorded runs informed the settings. Peak GPU memory was not logged.

After training, the adapters were merged into 16-bit weights, converted to GGUF with
llama.cpp\footnote{\url{https://github.com/ggml-org/llama.cpp}} and quantised to
Q4\_K\_M, a different 4-bit scheme from the NF4 used in training. All reported results
come from these exported files, served with llama-cpp-python (context 4,096 tokens, at
most 32 new tokens, temperature 0.1, repetition penalty 1.15) on an Apple M1 Pro laptop.
The hardware comes from the first author's records; the logs do not name it. We did not compare predictions before and after export, so the cost of the
export step itself is unknown.

\begin{table*}[width=\textwidth,cols=4,pos=t]
\caption{Training configuration of the learned models, as recorded in the training
scripts and logs. LLM losses are averaged over each epoch and cover all tokens of each
record; CodeBERT's log keeps no per-epoch average. The NF4 settings come from the
quantisation configuration of the base checkpoints.}
\label{tab:training}
\footnotesize\setlength{\tabcolsep}{4pt}
\begin{tabular*}{\tblwidth}{@{\extracolsep{\fill}} >{\raggedright\arraybackslash}p{2.6cm} >{\raggedright\arraybackslash}p{3.5cm} >{\raggedright\arraybackslash}p{4.2cm} >{\raggedright\arraybackslash}p{4.2cm} @{}}
\toprule
 & CodeBERT & Llama 3 8B (QLoRA) & Qwen 2.5 7B (QLoRA) \\
\midrule
Initial checkpoint & \texttt{microsoft/codebert-base} & \texttt{unsloth/llama-3-8b-bnb-4bit} & \texttt{unsloth/qwen2.5-7b-bnb-4bit} \\
Trainable / total parameters & 124,648,708 / 124,648,708 & 83,886,080 / 8,114,147,328 & 80,740,352 / 7,696,356,864 \\
Adaptation & full fine-tuning, classification head & \multicolumn{2}{>{\raggedright\arraybackslash}p{8.6cm}}{LoRA $r=32$, $\alpha=32$, dropout 0.05, on q, k, v, o, gate, up, down} \\
Precision in training & fp32 weights, fp16 mixed precision & \multicolumn{2}{>{\raggedright\arraybackslash}p{8.6cm}}{block projections frozen in NF4 with double quantisation (bitsandbytes); embeddings, norms and LM head in bf16; bf16 compute} \\
Objective & cross-entropy on 4 labels & \multicolumn{2}{>{\raggedright\arraybackslash}p{8.6cm}}{next-token cross-entropy on the whole sequence (prompt not masked)} \\
Optimiser, learning rate & AdamW, $2\times10^{-5}$, linear decay, no warm-up & \multicolumn{2}{>{\raggedright\arraybackslash}p{8.6cm}}{paged\_adamw\_8bit, $1\times10^{-4}$, cosine, 10\% warm-up} \\
Weight decay & 0.01 & \multicolumn{2}{>{\raggedright\arraybackslash}p{8.6cm}}{0.01} \\
Batch, epochs, steps & 16, 3, 3,000 & \multicolumn{2}{>{\raggedright\arraybackslash}p{8.6cm}}{8 ($2\times4$ accumulation), 3, 6,000} \\
Input limit & 512 tokens & \multicolumn{2}{>{\raggedright\arraybackslash}p{8.6cm}}{4,000 characters; 4096-token context} \\
Seed & 42 & \multicolumn{2}{>{\raggedright\arraybackslash}p{8.6cm}}{3407} \\
Hardware, software & Colab GPU (type not logged) & \multicolumn{2}{>{\raggedright\arraybackslash}p{8.6cm}}{A100-SXM4-40GB; Unsloth 2026.3.3, Transformers 5.2.0, PyTorch 2.10.0+cu128} \\
Training wall clock & 0:23:28 & 3:01:12 & 2:39:47 \\
Mean loss, epochs 1 / 2 / 3 & not logged per epoch & 1.091 / 0.915 / 0.760 & 1.054 / 0.916 / 0.795 \\
Deployed artefact & PyTorch weights (fp32) & \multicolumn{2}{>{\raggedright\arraybackslash}p{8.6cm}}{LoRA merged into 16-bit weights, GGUF Q4\_K\_M (llama.cpp)} \\
\bottomrule
\end{tabular*}

\end{table*}

\subsection{Evaluation protocol}
\label{sec:protocol}

\paragraph{Shared test reports.} Local LLM inference was slow, so the evaluation
scripts drew 2,000 of the 4,000 test reports with NumPy's generator seeded at 42 and ran
the four LLM configurations on that subset. We regenerated the sampling indices with the
same call, checked that every stored LLM prediction matches its test row in text and
gold label, and re-scored the six TF-IDF classifiers and CodeBERT from their saved model
files on the same 2,000 reports. The subset contains 495
Critical, 505 Major, 491 Minor and 509 Trivial reports. Re-scoring the non-LLM models
on all 4,000 reports reproduces the logged metrics to the fourth decimal
(Appendix~\ref{app:full}).

\paragraph{Parsing.} Zero-shot and fine-tuned outputs went through the same parser.
It tries to read the whole response, and then the span from its first \texttt{\{} to its
last \texttt{\}}, as a JSON object, and accepts the \texttt{severity} value if it equals
one of the four labels, ignoring case. Otherwise it strips punctuation from the whole
response and checks for the label names in the fixed order Critical, Major, Minor,
Trivial, taking the first one that occurs anywhere. Only when no label name occurs, as
in an empty response, a failed request or an answer such as ``Blocker'' or ``High'', does
it record Minor. Every report counts in the metrics; none is dropped. The raw responses
were not stored, so we cannot report how often each step fired or how many responses
broke the requested format. The zero-shot errors therefore mix wrong labels with
unusable output, and some Minor predictions may come from the default.

\paragraph{Metrics.} With $TP_c$, $FP_c$ and $FN_c$ the per-class counts, precision
and recall are $P_c=TP_c/(TP_c+FP_c)$ and $R_c=TP_c/(TP_c+FN_c)$, and the primary
metric is
\begin{equation}
  \text{macro-F}_1=\frac{1}{4}\sum_{c}\frac{2P_cR_c}{P_c+R_c}.
  \label{eq:f1}
\end{equation}
We also report accuracy, macro-averaged precision and recall, per-class precision and
recall, and confusion matrices.

\paragraph{Uncertainty and tests.} Intervals come from a percentile bootstrap over
reports~\citep{efron1993introduction}: we resampled the 2,000 reports with replacement
10,000 times and took the 2.5th and 97.5th percentiles of each statistic. The same
resamples are used for every model, so intervals for between-model differences are
paired. For ten comparisons selected from the research questions
(Table~\ref{tab:pairwise}) we report a two-sided bootstrap $p$-value for the macro-F1
difference, taken as twice the smaller share of resampled differences on either side of
zero (an approximation obtained by inverting the interval), and an exact McNemar test of
accuracy on the reports that exactly one of the two models gets
right~\citep{mcnemar1947note,dietterich1998approximate}. Each family of ten $p$-values is
corrected with Holm's procedure~\citep{holm1979simple}. We call a difference detectable
when its 95\% interval excludes zero and its adjusted $p$-value is below 0.05. Reports are
treated as independent, although the shared subset contains 18 repeated descriptions,
and each model was trained once, so seed-to-seed variation is not part of these
intervals.

\section{Results}
\label{sec:results}

\subsection{RQ1: fine-tuned LLMs against TF-IDF and CodeBERT}
\label{sec:res_overall}

\begin{figure}[pos=t]
\centering
\includegraphics[width=\linewidth]{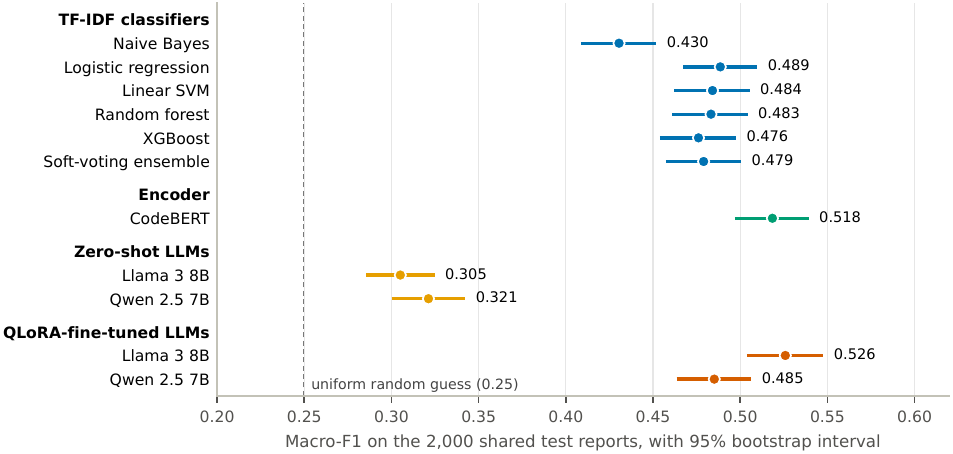}
\caption{Macro-F1 of all eleven configurations on the shared 2,000 reports, with
95\% bootstrap intervals.}
\label{fig:f1}
\end{figure}

Table~\ref{tab:main} and Figure~\ref{fig:f1} give the scores on the 2,000 shared
reports. They fall into three groups. The zero-shot LLMs sit at 0.305 and 0.321, well
below all six TF-IDF classifiers. Those classifiers span 0.430 to 0.489, with Naive
Bayes clearly last and the other five within 0.013 of each other; on these 2,000
reports their intervals overlap almost completely. CodeBERT (0.518) and fine-tuned
Llama~3 (0.526) form the top group, with fine-tuned Qwen~2.5 (0.485) close to the
better TF-IDF models.

\begin{table*}[width=\textwidth,cols=7,pos=t]
\caption{Results on the 2,000 shared test reports (495 Critical, 505 Major, 491 Minor,
509 Trivial). Acc.\ is accuracy; brackets give 95\% bootstrap intervals. The
last column (dedup.) repeats macro-F1 after removing the 159 shared reports whose text
also appears in the training set ($n=1{,}841$). Best macro-F1 in bold.}
\label{tab:main}
\small
\begin{tabular*}{\tblwidth}{@{\extracolsep{\fill}} llccccc @{}}
\toprule
Family & Model & Acc. & Macro-P & Macro-R & Macro-F1 [95\% CI] & F1, dedup. \\
\midrule
TF-IDF & Naive Bayes & 0.435 & 0.456 & 0.437 & 0.430 [0.409, 0.452] & 0.425 \\
 & Logistic regression & 0.491 & 0.488 & 0.490 & 0.489 [0.467, 0.510] & 0.474 \\
 & Linear SVM & 0.490 & 0.484 & 0.489 & 0.484 [0.462, 0.505] & 0.471 \\
 & Random forest & 0.491 & 0.486 & 0.491 & 0.483 [0.461, 0.504] & 0.455 \\
 & XGBoost & 0.479 & 0.475 & 0.479 & 0.476 [0.454, 0.498] & 0.445 \\
 & Soft-voting ensemble & 0.486 & 0.482 & 0.486 & 0.479 [0.457, 0.501] & 0.458 \\
\midrule
Encoder & CodeBERT & 0.518 & 0.522 & 0.517 & 0.518 [0.497, 0.539] & 0.508 \\
\midrule
Zero-shot & Llama 3 8B & 0.367 & 0.531 & 0.370 & 0.305 [0.286, 0.325] & 0.302 \\
 & Qwen 2.5 7B & 0.374 & 0.535 & 0.377 & 0.321 [0.300, 0.342] & 0.320 \\
\midrule
Fine-tuned & Llama 3 8B & 0.527 & 0.570 & 0.526 & \textbf{0.526} [0.504, 0.548] & 0.524 \\
 & Qwen 2.5 7B & 0.518 & 0.510 & 0.517 & 0.485 [0.463, 0.506] & 0.478 \\
\bottomrule
\end{tabular*}

\end{table*}

Table~\ref{tab:pairwise} tests the selected comparisons. Fine-tuned Llama~3 and
CodeBERT show no detectable difference: the macro-F1 difference is $+0.007$ with an interval from
$-0.017$ to $+0.032$, and of the 654 reports on which exactly one of them is right,
Llama gets 336 and CodeBERT 318. Fine-tuned Qwen~2.5 is 0.033 below CodeBERT on
macro-F1 (adjusted $p=0.031$), yet the two have the same accuracy (333 against 332
discordant reports). The same split appears between the two LLMs: Llama~3 leads Qwen~2.5
by 0.041 in macro-F1 while their accuracies do not differ. Macro-F1 separates these
models because it weights each class's F1 equally, and Qwen's errors are concentrated in
one class (Section~\ref{sec:res_class}). Against the soft-voting ensemble, fine-tuned
Llama~3 is ahead on both tests (macro-F1 $+0.047$, accuracy $+0.041$), CodeBERT is ahead
on macro-F1 ($+0.039$) but not on accuracy after Holm correction ($p=0.079$), and
fine-tuned Qwen~2.5 shows no detectable difference from it.

\begin{table*}[width=\textwidth,cols=5,pos=t]
\caption{Selected pairwise comparisons on the shared 2,000 reports. FT is fine-tuned,
ZS zero-shot, and the TF-IDF ensemble is the soft-voting ensemble. $\Delta$ is A minus
B. $b$ and $c$ count reports that only A or only B classifies correctly. Bootstrap
$p$-values cannot fall below $1/10{,}000$, hence the floor after Holm correction.}
\label{tab:pairwise}
\small
\begin{tabular*}{\tblwidth}{@{\extracolsep{\fill}} l c c c c @{}}
\toprule
Comparison (A vs.\ B) & $\Delta$ macro-F1 [95\% CI] & $p_{\text{Holm}}$, bootstrap & $b$ / $c$ & $p_{\text{Holm}}$, McNemar \\
\midrule
Llama 3 FT vs.\ CodeBERT & $+0.007$ [$-0.017$, $+0.032$] & 1.000 & 336 / 318 & 1.000 \\
Qwen 2.5 FT vs.\ CodeBERT & $-0.033$ [$-0.058$, $-0.008$] & 0.031 & 333 / 332 & 1.000 \\
Llama 3 FT vs.\ Qwen 2.5 FT & $+0.041$ [$+0.015$, $+0.066$] & 0.010 & 380 / 363 & 1.000 \\
Llama 3 FT vs.\ Llama 3 ZS & $+0.221$ [$+0.195$, $+0.247$] & $\le$0.001 & 581 / 262 & $<$0.001 \\
Qwen 2.5 FT vs.\ Qwen 2.5 ZS & $+0.164$ [$+0.133$, $+0.194$] & $\le$0.001 & 748 / 459 & $<$0.001 \\
CodeBERT vs.\ TF-IDF ensemble & $+0.039$ [$+0.014$, $+0.065$] & 0.010 & 373 / 309 & 0.079 \\
Llama 3 FT vs.\ TF-IDF ensemble & $+0.047$ [$+0.020$, $+0.074$] & 0.002 & 448 / 366 & 0.027 \\
Qwen 2.5 FT vs.\ TF-IDF ensemble & $+0.006$ [$-0.021$, $+0.033$] & 1.000 & 384 / 319 & 0.079 \\
Llama 3 ZS vs.\ TF-IDF ensemble & $-0.174$ [$-0.201$, $-0.147$] & $\le$0.001 & 385 / 622 & $<$0.001 \\
Qwen 2.5 ZS vs.\ TF-IDF ensemble & $-0.158$ [$-0.186$, $-0.130$] & $\le$0.001 & 367 / 591 & $<$0.001 \\
\bottomrule
\end{tabular*}

\end{table*}

For RQ1, then: on these reports fine-tuned Llama~3~8B shows no detectable difference
from fine-tuned CodeBERT and is ahead of the TF-IDF ensemble, while fine-tuned
Qwen~2.5~7B is below CodeBERT on macro-F1 and shows no detectable difference from the
ensemble. Neither LLM is ahead of the encoder.

\subsection{RQ2: zero-shot versus fine-tuned, per class}
\label{sec:res_class}

The fine-tuned runs score 0.221 (Llama~3) and 0.164 (Qwen~2.5) higher in macro-F1 than
the zero-shot runs (Table~\ref{tab:pairwise}). Part of that gap may come from the setup
and not from training: the zero-shot runs used instruction-tuned checkpoints inside
their chat templates, served by Ollama with a 150-token cap, while the fine-tuned runs
used base checkpoints on raw text, served by llama-cpp-python with a 32-token cap and a
repetition penalty; and zero-shot answers without a label name became Minor.
Section~\ref{sec:disc_ft} returns to this.

\begin{table*}[width=\textwidth,cols=10,pos=t]
\caption{Per-class precision (P) and recall (R) on the shared 2,000 reports, with a
95\% bootstrap interval for Critical recall.}
\label{tab:perclass}
\small
\begin{tabular*}{\tblwidth}{@{\extracolsep{\fill}} l cc cc cc cc c @{}}
\toprule
 & \multicolumn{2}{c}{Critical} & \multicolumn{2}{c}{Major} & \multicolumn{2}{c}{Minor} & \multicolumn{2}{c}{Trivial} & Critical recall \\
\cmidrule(lr){2-3}\cmidrule(lr){4-5}\cmidrule(lr){6-7}\cmidrule(lr){8-9}
Model & P & R & P & R & P & R & P & R & 95\% CI \\
\midrule
Soft-voting ensemble & 0.518 & 0.513 & 0.447 & 0.311 & 0.408 & 0.458 & 0.553 & 0.660 & [0.471, 0.557] \\
CodeBERT & 0.557 & 0.471 & 0.423 & 0.440 & 0.454 & 0.511 & 0.655 & 0.648 & [0.427, 0.515] \\
Llama 3 8B (zero-shot) & 0.741 & 0.162 & 0.388 & 0.450 & 0.319 & 0.829 & 0.677 & 0.041 & [0.129, 0.195] \\
Qwen 2.5 7B (zero-shot) & 0.700 & 0.127 & 0.396 & 0.366 & 0.322 & 0.886 & 0.722 & 0.128 & [0.099, 0.157] \\
Llama 3 8B (fine-tuned) & 0.632 & 0.358 & 0.412 & 0.707 & 0.517 & 0.493 & 0.720 & 0.546 & [0.316, 0.400] \\
Qwen 2.5 7B (fine-tuned) & 0.524 & 0.691 & 0.484 & 0.335 & 0.495 & 0.214 & 0.536 & 0.827 & [0.651, 0.731] \\
\bottomrule
\end{tabular*}

\end{table*}

\begin{figure*}[pos=t]
\centering
\includegraphics[width=.92\textwidth]{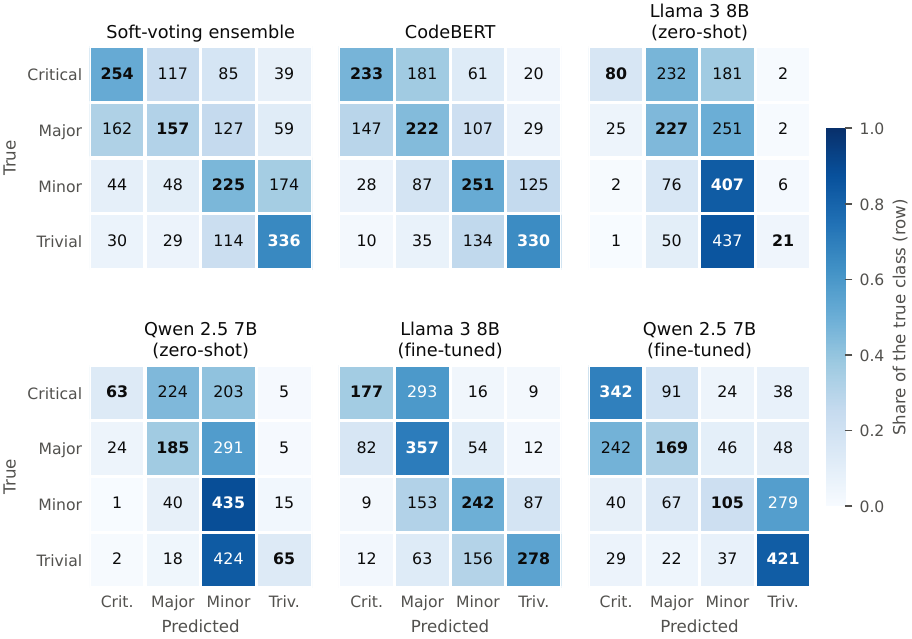}
\caption{Confusion matrices on the shared 2,000 reports for six configurations. Cells
show counts; shading shows the share of each true class (row). Rows are true labels,
columns predicted labels.}
\label{fig:confusion}
\end{figure*}

The per-class picture (Table~\ref{tab:perclass}, Figure~\ref{fig:confusion}) shows
how the two conditions differ. Both zero-shot runs pull reports towards Minor, possibly
in part through the parser default: Llama~3 predicts Minor for 1,276 of the 2,000
reports and Qwen~2.5 for 1,353, whereas Trivial is predicted 31 and 90 times. When they do say Critical they are usually right (precision
0.741 and 0.700), but they say it rarely, so Critical recall is 0.162 and 0.127.

After fine-tuning the two models settle into opposite trade-offs. Llama~3 leans
towards Major, which it predicts 866 times. It finds 177 of the 495 Critical reports
(recall 0.358, interval 0.316--0.400) and labels 293 of the rest Major, so 64\% of
Critical reports are missed, most of them one level down. Qwen~2.5 leans towards the
two ends of the scale. It finds 69\% of Critical reports (recall 0.691, interval
0.651--0.731) and 83\% of Trivial ones, at the price of calling 242 Major reports
Critical and 279 of the 491 Minor reports Trivial, leaving Minor recall at 0.214.
CodeBERT is more even, with recall between 0.440 and 0.648 for every class.

Most errors of every model fall between neighbouring levels: 70--73\% for the TF-IDF
classifiers, 81\% for CodeBERT, 79\% for fine-tuned Qwen~2.5 and 87\% for fine-tuned
Llama~3. For the soft-voting ensemble the largest confusions are Minor with Trivial
(288 reports) and Critical with Major (279), not Major with Minor (175). In short, for
RQ2, the fine-tuned runs use the whole scale where the zero-shot runs concentrate on Minor,
but the two fine-tuned models distribute their errors very differently, and neither
finds Critical reports reliably.

\subsection{RQ3: duplicates and class prevalence}
\label{sec:res_sensitivity}

Of the 2,000 shared reports, 159 have a description that also appears in the training
set. Dropping them (last column of Table~\ref{tab:main}) lowers the TF-IDF scores by
0.005 to 0.031, most for XGBoost and random forest, and CodeBERT's from 0.518 to 0.508.
The fine-tuned LLMs move by 0.007 or less, and the zero-shot runs, which never saw the
training set, by 0.003 or less. The ordering behind RQ1 is unchanged on the remaining
1,841 reports: fine-tuned Llama~3 (0.524) and CodeBERT (0.508) are 0.016 apart, and both
stay above every TF-IDF classifier (at most 0.474).

The small change for the fine-tuned LLMs does not show that they handle duplicates
better. On the 159 shared reports whose text and label appear in training, random
forest and XGBoost are right 86\% and 87\% of the time, CodeBERT 68\%, fine-tuned
Qwen~2.5 66\% and fine-tuned Llama~3 56\%; on the other shared reports the same five
models score 46\%, 44\%, 50\%, 51\% and 52\%. The gain on seen reports is 40 and 43
points for the two tree models, 18 for CodeBERT, 15 for Qwen~2.5 and only 4 for
Llama~3. With 159 reports each of these rates is uncertain by up to about $\pm$8 points, and
only 126 of the 159 were in the 12,800 reports the TF-IDF models were fit on. Fine-tuned
Qwen~2.5 therefore behaves much like CodeBERT, while fine-tuned Llama~3 gains little
from having seen a report. Three explanations are possible for Llama~3 and we cannot
separate them: limited fitting after three epochs of LoRA at a learning rate of
$10^{-4}$, the training loss, in which the 6 or 7 answer tokens are a small part of each
record, and the export from NF4 training to a Q4\_K\_M file. Since 175 of the 319
repeated test reports are Trivial, the effect of duplication falls mostly on that
class.

The test set is balanced, but the pool is not. As an estimate, not a measurement, we
re-weighted each model's confusion matrix to the pool's class mix (Major 48.6\%,
Critical 26.3\%, Minor 18.5\%, Trivial 6.6\%), assuming each model keeps its per-class
behaviour. Under that mix the estimated macro-F1 falls by 0.046 to 0.078 for the TF-IDF
classifiers, by 0.056 for CodeBERT (to 0.462) and by 0.080 for fine-tuned Qwen~2.5 (to
0.405), but by only 0.009 for fine-tuned Llama~3 (to 0.517). The reason is Trivial
precision. With Trivial reports scarce, models that predict Trivial generously produce
mostly false alarms: Trivial precision drops to 0.242 for Qwen~2.5 and 0.333 for
CodeBERT, against 0.424 for Llama~3, whose errors go to the now common Major class.
Under this estimate fine-tuned Llama~3 leads CodeBERT by 0.055 (95\% bootstrap interval
0.029 to 0.081), but the interval does not cover the re-weighting assumption itself.
For RQ3, then, duplication inflates the TF-IDF scores somewhat and leaves the LLM
comparison intact, while the balanced test set may hide an advantage of fine-tuned
Llama~3 over CodeBERT under a Major-heavy stream; only a test set drawn at natural
prevalence can confirm or reject it.

\subsection{Inference and training cost}
\label{sec:res_cost}

Table~\ref{tab:latency} lists the wall-clock times logged by the evaluation scripts on
the M1 Pro. CodeBERT needs 49\,ms per report and fine-tuned Llama~3 1.31\,s, about 27
times longer, with no detectable difference in macro-F1 on balanced data. The
fine-tuned runs were 4.5\% (Llama) and 12.5\% (Qwen) faster than the zero-shot ones.
We do not credit this to fine-tuning: the loops differ in serving runtime and output
cap, and the zero-shot loop also pauses 50\,ms per request, which adds 100\,s per run
(3.6\% of the Llama zero-shot time). Training took 23 minutes for CodeBERT, and 3 hours (Llama~3) and 2 hours
40 minutes (Qwen~2.5) for QLoRA, on the A100 (Table~\ref{tab:training}).

\begin{table}[width=\linewidth,cols=5,pos=t]
\caption{Inference wall-clock time on the Apple M1 Pro, from the most recent logged
evaluation runs. The LLM loops also include prompt building and parsing; in the
CodeBERT and LLM loops the first call also loads the model.}
\label{tab:latency}
\small
\begin{tabular*}{\tblwidth}{@{\extracolsep{\fill}} lrrr>{\raggedright\arraybackslash}p{4.6cm} @{}}
\toprule
Model & Reports & Total (s) & Per report & Timed loop covers \\
\midrule
TF-IDF classifiers (six) & 4,000 & 0.39--1.09 & 0.10--0.27 ms & load model, vectorise, predict \\
CodeBERT & 4,000 & 196.3 & 49 ms & per-report tokenise and forward pass \\
Llama 3 8B (zero-shot) & 2,000 & 2,742 & 1.37 s & HTTP call to local Ollama, 150-token cap, 50 ms pause \\
Qwen 2.5 7B (zero-shot) & 2,000 & 3,076 & 1.54 s & HTTP call to local Ollama, 150-token cap, 50 ms pause \\
Llama 3 8B (fine-tuned) & 2,000 & 2,619 & 1.31 s & llama-cpp-python call, 32-token cap \\
Qwen 2.5 7B (fine-tuned) & 2,000 & 2,693 & 1.35 s & llama-cpp-python call, 32-token cap \\
\bottomrule
\end{tabular*}

\end{table}

\section{CyberTriage: serving the classifier}
\label{sec:system}

The fine-tuned models are packaged in CyberTriage, an open web application with a
React front end and a FastAPI back end, released with the analysis code at
\coderepo. It runs in the browser, so practitioners can
use the models without running scripts, and, like our earlier web-based platform with
multimodal search~\citep{bayan2025enhanced}, it accepts more than one kind of input:
reports can be typed, uploaded as PDF, DOCX, TXT, LOG or JSON files, or dictated through
the browser's speech-recognition interface. A registry lets the user pick any of the
evaluated models. Analysed
reports are kept in a history view and can be exported as a Jira import file, with
Critical, Major, Minor and Trivial mapped to the Jira priorities Highest, High, Low and
Lowest.

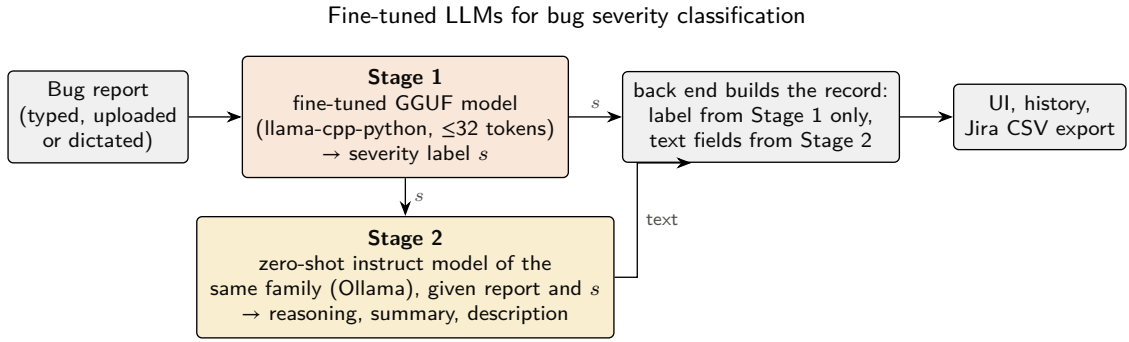
\begin{figure}[pos=t]
\centering
\begin{tikzpicture}[font=\sffamily\footnotesize, node distance=5mm and 7mm,
  box/.style={draw=black, line width=0.5pt, fill=gray!12, rounded corners=2pt, align=center, inner sep=4pt, text=inkA},
  stage/.style={draw=black, line width=0.5pt, fill=white, rounded corners=2pt, align=center, inner sep=5pt, text=inkA},
  arr/.style={-{Stealth[length=2mm]}, draw=black, line width=0.5pt}]
  \node[box] (in) {Bug report\\(typed, uploaded\\or dictated)};
  \node[stage, fill=famFT!15, right=of in] (s1) {\textbf{Stage 1}\\fine-tuned GGUF model\\(llama-cpp-python, $\le$32 tokens)\\$\rightarrow$ severity label $\hat{s}$};
  \node[stage, fill=famZS!18, below=of s1] (s2) {\textbf{Stage 2}\\zero-shot instruct model of the\\same family (Ollama), given report and $\hat{s}$\\$\rightarrow$ reasoning, summary, description};
  \node[box, right=of s1] (merge) {back end builds the record:\\label from Stage 1 only,\\text fields from Stage 2};
  \node[box, right=of merge] (out) {UI, history,\\Jira CSV export};
  \draw[arr] (in) -- (s1);
  \draw[arr] (s1) -- node[right, font=\sffamily\scriptsize, text=inkB]{$\hat{s}$} (s2);
  \draw[arr] (s1) -- node[above, font=\sffamily\scriptsize, text=inkB]{$\hat{s}$} (merge);
  \draw[arr] (s2.east) -- ++(3mm,0) |- node[pos=0.25, right, font=\sffamily\scriptsize, text=inkB]{text} ($(merge.south west)!0.25!(merge.south east)$);
  \draw[arr] (merge) -- (out);
\end{tikzpicture}
\caption{Two-stage analysis in CyberTriage. Any severity the Stage~2 model writes is
discarded by the back end; the stored label is always Stage~1's.}
\label{fig:pipeline}
\end{figure}

For the fine-tuned models, a single analysis runs in two stages (Figure~\ref{fig:pipeline}).
Stage~1 is the fine-tuned classifier evaluated above and produces only the label. Stage~2
sends the report and that label to the zero-shot instruction model of the same family.
Its prompt states the label, tells the model not to disagree with it, and asks for a
technical justification, a one-line summary and a short description as JSON
(temperature 0.3, up to 1,024 tokens). The back end copies only those three fields from
the Stage~2 response; a severity field in that response, if present, is ignored. The
guarantee is therefore narrow and enforced in code, not by the prompt or the interface:
Stage~2 cannot change the stored label. It is not a guarantee of correctness. The
pipeline's label is Stage~1's label, which is wrong for 47\% (Llama~3) and 48\%
(Qwen~2.5) of the shared test reports, and the Stage~2 text is a post-hoc rationale
written by a different model from the one that decided. We expect it to justify a wrong
label as readily as a right one, but we have not evaluated the explanations, and we did
not time Stage~2, so we give no end-to-end latency for the two-stage path.

\section{Discussion}
\label{sec:discussion}

\subsection{What fine-tuning bought, and what it did not}
\label{sec:disc_ft}

The gap between the zero-shot and fine-tuned runs is large, but it should not be read
as the value of task learning alone. Training was not the only thing that changed
(Section~\ref{sec:res_class}): checkpoint, prompt template, serving stack, quantisation
format and decoding settings changed with it. With the raw responses gone we also
cannot separate format failures from wrong answers, so the Minor bias of the zero-shot
runs may partly reflect the parser default. A clean test of what fine-tuning adds would
run the same checkpoint under four conditions (plain zero-shot, zero-shot with output
constrained to the four labels, constrained few-shot, and fine-tuned with the same
constraint) and report output validity separately. We have not done this.

What the data do show is that fine-tuning on 16,000 labelled reports did not take an 8B
model detectably beyond a fully fine-tuned 125M encoder on this test set; the interval
allows a difference of up to 0.032 in favour of Llama~3 or 0.017 in favour of CodeBERT.
On the balanced test set the practical case for the LLM is therefore weak: it costs far more per report on the same
laptop (Table~\ref{tab:latency}) and three hours of A100 time to train. Our prevalence
estimate is the one result pointing the other way, and it depends on an assumption we
could not test.

The two fine-tuned models also disagree with each other more than their aggregate
scores suggest. Under identical data, prompt and hyperparameters they ended with
opposite biases, one towards Major and one towards the ends of the scale. We trained each model once and cannot say how much of
this would survive another seed, so we do not attribute it to the architectures.

\subsection{Implications for triage}

For triage, the per-class numbers matter more than the aggregate. Fine-tuned Llama~3
misses 64\% of Critical reports and Qwen~2.5 misses 31\%; CodeBERT misses 53\%. With
Critical recall between 0.36 and 0.69 for these three models, and lower for the zero-shot
runs, no model here should route reports without a person checking the result, least of all where a missed Critical report has a real
cost. The models are more useful as a second opinion than as a gate. Because most
misses land one level down, reviewing every report predicted Major or above catches
most Critical reports, at the cost of a long queue: on our balanced set this covers
95\% of Critical reports for fine-tuned Llama~3 but means reviewing 57\% of all
reports, against 84\% and 47\% for CodeBERT. At the pool's class mix, where Major is
common, the queue would grow to about 74\% and 62\% of reports. With score-based models one could instead
lower the decision threshold for Critical and choose it on held-out data; we did not
evaluate this.

\subsection{Threats to validity}
\label{sec:threats}

We follow the usual four categories~\citep{wohlin2012experimentation}.

\textit{Construct validity.} The ground truth is the severity recorded in Bugzilla,
which is noisy~\citep{tian2016unreliability}; 41 identical descriptions in our pool
carry conflicting labels, and we did not audit a sample by hand. Merging
\textit{blocker} into Critical and dropping \textit{normal} define a narrower task than
real triage, where \textit{normal} is the most common label. Macro-F1 on a balanced set
rewards equal performance across classes, which is not the same as the cost of errors in
practice.

\textit{Internal validity.} The zero-shot and fine-tuned conditions differ in more than
training (Section~\ref{sec:disc_ft}). The LLM outputs were parsed with a fallback whose
firing rate is unknown. CodeBERT's checkpoint was chosen on the test split, although
the choice coincided with the last epoch. The TF-IDF classifiers saw 12,800 training
reports and the other models 16,000. Each model was trained once with a single seed,
and LLM inference sampled at temperature 0.1 without a fixed seed, so run-to-run
variation is not covered by our intervals. We did not compare the NF4 training-time
models with their Q4\_K\_M exports, and the Ollama version, model digests and context
setting of the zero-shot runs were not recorded.

\textit{External validity.} All data come from one project, Eclipse, and from reports
that were later resolved; about 1\% of descriptions include notes added after filing,
and labels may have been revised after filing. The test set is balanced, the
sample includes repeated reports, and we have neither a natural-prevalence test set nor
a temporal or cross-project split. The prevalence estimate in
Section~\ref{sec:res_sensitivity} is a re-weighting and ignores the \textit{normal}
reports that would dominate a real stream.

\textit{Conclusion validity.} The LLMs were evaluated on 2,000 of the 4,000 test
reports, which widens intervals; differences below about 0.03 in macro-F1 are not
resolvable here. The bootstrap treats reports as independent, although 41 test rows
repeat a bug ID already in the test split (18 repeated descriptions in the shared
subset). We limited formal testing to ten comparisons selected from the research
questions and corrected for multiplicity within each test.

\section{Conclusion}
\label{sec:conclusion}

We compared TF-IDF classifiers, a fine-tuned CodeBERT encoder and two open 7--8B LLMs,
prompted and QLoRA-fine-tuned, on the same 2,000 Eclipse bug reports, using numbers
rebuilt from stored predictions. The zero-shot runs scored below every TF-IDF
classifier. The fine-tuned runs scored far higher, in a setup that changed more than
training. Fine-tuned Llama~3~8B showed no detectable difference from CodeBERT, at a much
higher cost per report, and fine-tuned Qwen~2.5~7B was below CodeBERT on macro-F1. The
fine-tuned models differed most in how they traded one class against another, and none
of them found Critical reports reliably enough to work without review. The data audit
showed that the public dump is Eclipse-only and repeats rows, and that the repeats reach
the test set and inflate the TF-IDF scores while leaving the LLM comparison as it was.

The next steps follow from the gaps listed above: a natural-prevalence test set and a
temporal split, deduplication before sampling, a validation split for every tuning
decision, several seeds per model, a general-purpose encoder such as RoBERTa or
DeBERTa as a second baseline, and the four-condition comparison of prompting and
fine-tuning with constrained output. The split identifiers, per-report predictions and the script that produces every table and figure in this paper are released so that
these extensions can start from the same records.

\appendix
\section{Re-scored baselines on the full test split}
\label{app:full}

Table~\ref{tab:full} gives the TF-IDF classifiers and CodeBERT on all 4,000 test
reports, re-scored from their saved model files, next to the macro-F1 and wall-clock
time of the most recent logged evaluation run. The re-scored values match the logged
ones.
Scores on the full split are 0.002 to 0.016 higher than on the 2,000-report subset, so
the subset is a slightly harder draw for these models.

\begin{table}[width=\linewidth,cols=7,pos=h]
\caption{TF-IDF classifiers and CodeBERT on the full 4,000-report test split.}
\label{tab:full}
\small
\begin{tabular*}{\tblwidth}{@{\extracolsep{\fill}} lcccccc @{}}
\toprule
Model & Accuracy & Macro-P & Macro-R & Macro-F1 & Logged F1 & Logged time (s) \\
\midrule
Naive Bayes & 0.450 & 0.472 & 0.450 & 0.444 & 0.4439 & 0.52 \\
Logistic regression & 0.493 & 0.489 & 0.493 & 0.491 & 0.4908 & 0.39 \\
Linear SVM & 0.495 & 0.488 & 0.495 & 0.489 & 0.4887 & 0.60 \\
Random forest & 0.501 & 0.495 & 0.501 & 0.494 & 0.4939 & 1.09 \\
XGBoost & 0.495 & 0.490 & 0.495 & 0.492 & 0.4917 & 0.58 \\
Soft-voting ensemble & 0.498 & 0.492 & 0.498 & 0.490 & 0.4904 & 1.02 \\
CodeBERT & 0.531 & 0.535 & 0.531 & 0.532 & 0.5322 & 196.27 \\
\bottomrule
\end{tabular*}

\end{table}

\printcredits

\section*{Declaration of competing interest}
The authors declare no competing interests.

\section*{Declaration of generative AI and AI-assisted technologies in the writing process}
During the development of the CyberTriage software the first author used AI coding
assistants (GitHub Copilot and Cursor). During the preparation of this manuscript the
authors used Claude (Anthropic) to help re-run the verification analysis and to edit the
text. The authors reviewed and edited all content and take full responsibility for the
publication.

\section*{Data availability}
The source data are public at
\url{https://huggingface.co/datasets/AndressaStefany/bug-reports}. The repository
\coderepo{} holds the code, the training notebooks
with their logs, the logged evaluation metrics, the split identifiers (bug ID, split,
label, membership in the shared subset, duplicate flag), the per-report predicted labels
of all eleven configurations, and the analysis script, which rebuilds the split from the
source data and recomputes every table and figure from these files. The trained model
files are not redistributed; every reported number can be recomputed from the released
per-report predictions. The raw LLM responses were not stored.

\section*{Acknowledgements}
This work started from the first author's master's thesis at Nazarbayev University (2026),
whose original code is at \url{https://github.com/asset8k/bug-triage-app}. The first author
and the co-authors wrote the first draft together. After graduating, the first author gave
his written consent for the co-authors to continue the work: its revision, new experiments,
methods and findings, and journal submission and revision.
We also thank Professor Siamac Fazli for recommendations on revisions.

\bibliographystyle{elsarticle-num-names}
\bibliography{references}

\end{document}